# Unity, Disunity and Pluralism in Science

David A. Edwards and Stephen Wilcox
(1980)

**ABSTRACT**

We discuss the problems of consensus and unity in science. The natural sciences seem to contrast with many other areas of endeavor in that a high level of consensus seems to exist in them. However, a careful analysis of the structure of particular physical theories, such as those concerned with electrons, shows that there is great heterogeneity of both theory and methodology. We argue that the natural science community tends to be tolerant of diversity. We contrast this tolerance among natural scientists with the more overt disagreement occurring in the social sciences and humanities. Our central theme is that many intellectual problems arise from straining too hard to make a successful perspective into a total worldview.

Nothing comes to pass in nature, which can be set down to a flaw therein: for nature is always the same, and everywhere one and the same in her efficacy and power of action; that is, nature's laws and ordinances, whereby all things come to pass and change from one form to another, are everywhere and always the same; so that there should be one and the same method of understanding the nature of all things whatsoever, namely, through nature's universal laws and rules – SPINOZA

1. **Introduction**

The revised modern scientific version of Genesis resembles in many ways the older version. The following figure compactly summarizes modern cosmology. All that exists is assumed to fit together into a compact hierarchy, all of which is ultimately reducible to the lower levels, which are both ontologically more fundamental and older in time than the upper levels. The background represents space-time. Time itself only goes back a finite amount usually estimated as around 13.7 billion years.

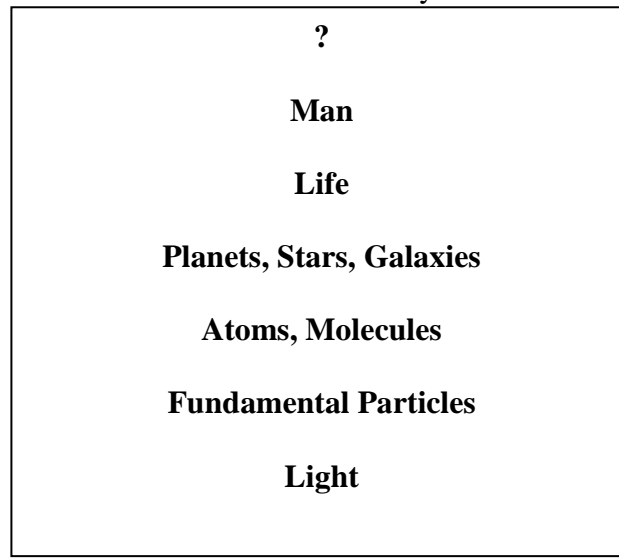

Although this presents a nice picture, all does not fit together quite so neatly. Einstein, the founder of modern cosmology, hoped to be able to explain everything by a purely geometric theory of curved space-time. Complex patterns of bumps and ripples in space-time were to be the underlying reality for what we usually call light, matter, animals and people. His program had reasonable success with bulk light and matter. For instance he was able to obtain the Newtonian Gravitational Theory as a first approximation, and higher order approximations resolved the anomaly in the perehellion shift of the orbit of Mercury (a major open problem in celestial mechanics for over fifty years). But his approach was never able to incorporate the modern theory of photons and electrons. His attempt to entirely bypass modern quantum theory was never fully successful. The current approach is simply to add quantum phenomena to space-time. More precisely, one assumes that in each region of space-time one can perform certain experiments such as applying photon detectors, electron detectors, etc. Each region has a disposition for responding to the experiments in various ways. Thus, in our figure the early region marked "Light" is to be understood as a region having a disposition to respond affirmatively to photon detectors but negatively to electron detectors. In other words, the elementary particles of modern quantum mechanics are not things, in the ordinary sense, but more akin to glimpses of things which cannot be said to exist when no one is looking. This distinction is made necessary by the attribution of stochastic properties to the independent existents. Thus, what does exist when we are not looking is a probabilistic disposition. Therefore, it is not correct to say that the early universe consisted of photons, although we might say that it consisted of a particular type of dispositional structure. According to the proponents of this quantum field theory, it is supposed to encompass all phenomena. It is having reasonable success in predicting such phenomena as the matter-anti-matter distribution and the hydrogen-helium distribution in the universe at various times. But what of biological phenomena? Obviously, quantum field theory is not presently adequate for predicting biological phenomena. The problem goes far deeper than mere lack of prediction, however. From what we have said so far it should be clear that it would be a mistake to say that a tree, for example, is "made up" of fundamental particles. It would follow that trees do not exist unless they are looked at, and this is simply not what we mean by tree ( the rather silly claims of the radical skeptic, notwithstanding ). On the other hand, it does not seem much better to say that a tree is really a probabilistic dispositional structure.

Thus there seems to be an irreconcilable incompatibility between the language of biology and that of quantum mechanics similar to the situation holding between quantum mechanics and general relativity. It is not just that the reduction of biology to physics has not been worked out. Rather it would appear that such a reduction is impossible in principle. As another example, a geologist wants to say that India collided with Asia, throwing up the Himalayas. He does not want to use probabilistic language. For him this event is a fact which he can almost see. The physicists language becomes even more inappropriate when applied to history or to Wigner's friend. If the reader is still unconvinced about the problems of reconciling the use of dispositional language with the points of view of biology, geology, and history, and feels that he can satisfactorily follow  J.S. Mill in considering objects as permanent possibilities of experience, then we invite him to contemplate the following recent developments at the forefront of physics. We have described the big bang model as a quantum field theory in a curved space-time. But J. Wheeler has long advocated a much more ambitious program of quantizing geometry itself (Wheeler, 1964, 1968). This program is now moving from the stage of being a philosophical vision to a well developed mathematical theory (see the surveys of DeWitt, 1979, and Hawking, 1979). Soon there may be strong experimental confirmation, for instance, in the scattering of light past black holes and in the properties of the background radiation due to conditions in the very early universe ($10^{-43}$ second after the beginning ). In these theories **the past did not actually occur**. Instead one has a superposition over the ensemble of all pasts. No particular past really occurred, but they were all virtually present. And this virtual presence shows up in interference terms occurring in the scattering matrix. The philosophical implications of this
theory were explored in the fifties by Everett (see DeWi t t and Graham, 1973). We don't believe that there is any possibility of reconciliation between this viewpoint and ordinary common sense, not to mention the rest of science. Even Carl Sagan would
be too embarrassed to seriously present this worldview to the public. The traditional conception of the unity of science is belied by these examples. The various sciences do not fit together into a single coherent picture. In fact, we will argue below that incompatibilities are commonplace even within a single science and often even within a single scientific paper. Moreover, these incompatibilities are often not to be viewed as defects needing remedy, but, instead, as rational responses to a very complex situation. Indeed we argue below that the Euclidean ideal of unity must give way to that of a more sophisticated conception.

In #2 we survey some of the inevitable problems generated by the attempt to build a thoroughgoing Euclidean world view. In #3 we discuss three episodes in science which epitomize the eternal dialectic between Euclidean and non-Euclidean approaches
to the Universe. In #4 we present a viewpoint we call perspective realism. In #5 we show that non-Euclidean viewpoints even pervade the exact sciences. In #6 we point out that modern biology refuses to take the ontology of modern physics (i.e., quantum mechanics) seriously and the changes of language that would be required of biology to so adapt to quantum theory. We also give a long quote of Pauling's which clearly shows chemistry's independence of physics. In #7 we consider the relationship between the humanities and sciences and in #8 we reconsider the possibilities open to theology. In #9 we draw our conclusions.

## 2. Grand Visions and Their Implications

There have been several bold attempts in the last few millenia to give a simple answer to the question, "What is the stuff of the universe?" The early Greek philosophers offer us several grand visions . Thales tells us that all is water, Anaximenes that all is air, Heraclitus that all is fire. The view of Parmenides that all is being is particularly intriguing, as is the claim of Pythagoras that all is number. And, of course, there is the atomism of Lucretius and Democritus which has much in common with more modern conceptions. By comparison, Aristotle's ontology is much more complex.

For Aristotle, the universe is made up of a heterogeneous nested hierarchy of locations, each of which contains substances to which are associated various qualities. Some of the qualities are realized, some are unrealized. The two great cosmologies which dominate our own era are those of Newton and Einstein. Newton's basic elements are particles which have very few properties (e.g., size, mass, attractive force) and exist in a homogeneous, three-dimensional s pa c e and one-dimensional time framework. And, as we have seen, Einstein's universe is as spare as any of the early Greeks'. These grand reductionist programs are things of great beauty, and some (e.g., Newton 's and Einstein's ) have displayed remarkable predictive power. What they all have in common, however, is that they leave no place for other viewpoints such as our ordinary common-sense view of the world. Since the world is said to be "really nothing but --, " our tables and chairs, theaters and ethical theories are robbed of ontological status. And this is the case even though the actual reduction of these entities to the elements of the cosmological theory in question is almost never accomplished. Thus, we (qua ordinary human beings) have been continually reminded throughout the centuries that the realm we inhabit is a mere fiction. We are condemned to Plato's cave having to make do with shadows on the wall. There is something profoundly dissatisfying about this. In fact, as we will show below, it is more than just dissatisfying.

Locke's epistemology is the classic articulation of the conflict between the viewpoint of the ordinary person and that of the physicist. For Locke, knowledge comes from two sources--the physical world and our own mind. The former is represented by the "primary" qualities, the latter by the "secondary." It is no coincidence that Locke was a contemporary of Newton. The primary qualities include solidity, extension, figure (i.e., shape), motion or rest, and number. These qualities are in external objects. They are real. The secondary qualities which include all other properties of objects (e.g., color, heat, taste, etc.) are in the beholder rather than the external world.
Locke tells us that "whatever reality we by mistake attribute to them [i.e., the secondary qualities], are in truth nothing in the objects themselves." (Locke, 1690, Book II, VII I.14). Although Locke's primary qualities are not identical to those which form the elmental properties within Newton's system, the influence is clear. For example, in describing how the secondary qualities are derived from the primary ones,
Locke explicitly invokes a Newtonian explanation: After the same manner that the ideas of these original qualities are produced in us, we may conceive that the ideas of secondary qualities are also produced, vis. by the operation of insensible particles on our senses. For, it being manifest that there are bodies and good store of bodies, each whereof are so small, that we cannot by any of our senses discover either their bulk, figure, or motion  let us suppose at present that the different motions and figures, bulk and number, of such particles, affecting the several organs of our senses, produce in us those different sensations which we have from the colors and smells of bodies (book II , VII I .13 ). Thus, Locke gave favored onto logical status to a certain viewpoint (i.e., that of  Newtonian physics), and inevitably had to relegate all other viewpoints to a questionable status. His titles to sections VII I.l7 and VIII I S of Book II tell the whole story: "The ideas of the primary alone really exist." "The secondary exist in things only as modes of the primary." And as the night follows the day, a dualism like Locke's leads either to a radical idealism or to a radical skepticism. Hence, Berkeley and Hume followed upon the heels of Locke.

What Berkeley showed, basically, was that all of the arguments which Locke used to show that some qualities are secondary apply as well to the primary qualities. He tells us let anyone consider those arguments which are thought manifestly to prove that colours and taste exist only in the mind, and he shall find they may with equal force be brought to prove the same thing of extension, figure, and motion (1710, 10). If some observed properties are in the world and others merely in our head, how can we distinguish the two? Or more to the point, how can we claim that anything exists at al1 except what is in our head? In other words, Locke's admission that we live partially in a world of appearance makes it difficult to maintain the qualifier partially. As Berkeley (1710) puts it:

Again, I ask whether those supposed originals or
external things, of which our ideas are pictures or
representations be themselves perceivable or not?
If they are, then they are ideas and we have gained
our point; but if you say they are not, I appeal to
anyone whether it be sense to assert a colour is
like something which is invisible (8 ) .

Later in the treatise Berkeley explicitly invokes physics.
He uses it as a stick to beat the realist into submission:

Matter, I say, and each particle thereof, is according
to them infinite and shapeless, and it is the mind that
frames all that variety of bodies which compose the
visible world, any one where of does not exist longer than
it is perceived (47).

Of course, Berkeley, by invoking God, ultimately opted for a position quite similar to Locke's, the mind of God playing the role that the external world played for Locke. It remained for Hume to remove God from the picture and thus end up with radical skepticism. Hume tells us that "It must certainly be allowed that nature has kept us at a great distance from all her secrets, and has afforded us only the objects of a few superficial qualities of objects, while she conceals from us those powers and principles on which the influence of those objects entirely depends (1739, 29)."

For Hume, all facts reduce to cause and effect, and cause and effect reduces to the conclusion from experience that the future will be like the past. But this conclusion is nothing but an unfounded leap. Therefore all knowledge is founded upon a very shaky foundation. We many not even be justified in claiming that knowledge is possible.

Locke, Berkeley and Hume mapped out the territory in which the philosophical debates of the next two hundred years took place. The British empiricists generally tried to salvage a dualism similar to Locke's, whereas the German idealists tended to follow Berkeley and Hume in looking askance at the primary qualities. Hence, we have, on the one hand, James and John Stuart Mill, whose elaborate associationisms where essentially schemes to get from primary to secondary qualities, and on the other hand Kant and Fichte. Kant pushed the primary qualities into
the realm of the unknowable. Fichte did away with them altogether.

In summary, the attempt to account for all that exists within one grand scheme inevitably gives birth to a skepticism which calls into question even the grand scheme itself. This is particularly problematical when a reductionist program is widely accepted, and even more so when it is widely accepted but left unexamined. In our opinion, that is exactly the situation we have today. The great success of physics has led to its near universal acceptance as the description of what is real. This is not to say that the average person knows much about physics, or that his conception of the physical really bears much resemblance to the physicist's. On the contrary, the typical conception of the
"real" physical world contains a hodgepodge of concepts f rom Aristotle, Newton, Einstein and commonsense. This conception does not, however, get one very far in most realms of human inquiry. Therefore, despite the work of modern philosopher's such as Wittgenstein, Ryle, and Austin, variations on the epistemologies of Locke, Hume, and Berkeley are ubiquitous. They pervade modern psychology and linguistics, art and literary criticism, and have even made in roads in biology and the philosophy of science. The arguments are surprisingly similar in the various fields. They typically begin with the demonstration that the physicist's description is inadequate. This leads to the proposition that the phenomenon in question forms an autonomous system which may or may not be related to reality. Thus, brains are about brains, minds about minds, art is about art, language is about language, etc. Sometimes there is an added twist in that the attempt is made to reduce everything to the phenomenon under study.

Psychology is an interesting case in point. From its beginnings, psychology has followed Locke in making a distinction between primary and secondary qualities, with the latter assumed to be its major province. The picture that Helmholtz (1867) drew was of the mind divided into sensations (described by the physicist) on the one hand and unconscious inferences on the other. The latter were designed to fill the gap between the physicist's world and the ordinary perceiver's. In other words, the perceiver begins with meaningless patterns of energy, and infers tables, chairs, etc., from them.

The jargon has, of course, changed over the decades. Today psychologists speak of sensory input and stages of information processing, for example, but the basic models are indistinguishable from Helmholtz's, which is to say from Locke's. The
common use of the term "constructivism" within psychology reveals
the Lockean spirit. Since the "input" is impoverished the person must construct his or her world out of the available building blocks. For example, Gregory (1966) desc ribes the problem of visual perception thus:

Given the slenderest clues to the nature of surrounding objects we identify them and act not so much according to what is directly sensed, but to what is believed (p.I I).

And just as Locke led to Hume, radical skepticism is unavoidable in modern psychology. Suppose our inferences are incorrect. Since all we have are our own representations, how do we know there is a world at all? This is just the train of thought that Fodor (1980) has recently presented in the context of arguing for what he calls "methodological solipsism." Fodor tells us that "a naturalistic psychology, i.e., a psychology which explicates the organism's relationship to the world, is, for practical purposes, out of the question" (p. 70). A naturalistic psychology would require "that the stimulus be physically specified" (p.71). What does Fodor mean by physical? He makes it clear when he points out that "If they [i.e., naturalistic psychologists] really had to wait for the physicists to determine the description how would the psychology get off the ground?" (p.71). Thus Fodor joins his forbears in deferring to the physicist for the description of what it means to be physical, and as a consequence, finding himself left with solipsism.

The biological version of this is represented in particularly blunt form by Maturana and Varela (1980). According to them, living systems are characterized by autonomy. The following depiction of their approach indicates that they do not mean autonomy in a trivial sense:

In other words, the new approach required us to treat seriously the activity of the nervous system as determined by the nervous system itself, and not by the external world, thus the external world would only have a triggering role in the release of the internally determined activity of the nervous system (p.XV).

In a discussion of how they arrived at their position, they refer to the discovery that one had to close off the nervous system to account for its operation, and that perception should not be viewed as a grasping of an external reality, but rather as the specification of one (p. XV). How did they discover this? In doing brain research, they found that they could not account for the manifold chromatic experiences of the observer by mapping the visible colorful world upon the activity of the nervous system" (p. XIV). In other words, they began with a characterization of the physical input to the system, and when this characterization did not allow them to predict cell firings, rather than attempting a reconceptualization of the inputs, they concluded that the nervous system creates its own reality. We find the same type of thinking in linguistics as represented by the rejection of referential accounts of semantics. Frege's (1960) account of meaning has a very Lockean flavor.

Meaning at the word level is a combination of reference which is related directly to the world and sense which is not. His classic example was that Venus is referred to both as the morning star and as the evening star. Both of these terms have the same referent so their meaning must differ in some other way, i.e., they must have different senses. Again we find a conflation of the scientist's viewpoint with the ordinary person's.

The two terms in question must refer to the same thing because the astronomer tells us they are the same. At the sentence level, Frege went even further: "We are therefore driven into accepting the truth value of a sentence as constituting its reference" (p, 63). It is a short step from Frege's position to the view that language is a completely autonomous system. Since language does not refer to the world as defined by the physicist, it must not refer to any thing, except possibly concepts (which just pushes the problem back one step) or language itself. In other words language is thought to be an "autonomous" system just like the living organism is for Maturana and Varela and the mind is for Fodor. Indeed, Fodor (1980) invoked Bloomfield's claim that the pursuit of semantics is basically impossible. The argument goes that since, for example, salt really refers to NaCl , we would have to rely upon chemistry in order to give a coherent account of minerals. It follows that we cannot say much about what words refer to because the appropriate scientific reductions have not been performed.

In modern linguistics, reference is often ignored altogether. Since language is treated as an autonomous system, the role of semantics becomes one of accounting for phenomena such as synonymy, ambiguity, and homonymy which can be treated as relations between sentences or between words, rather than as relations between language and the world. Treating language in this way has some interesting implications. For one thing it puts a strange cast upon linguistic analysis, which dominates British and American philosophy. If the task of philosophy is the analysis and reformulation of language, and if language does not refer to the world, the mainstream of modern philosophy begins to seem like a solipsistic endeavor. Does philosophy become the study of philosophical questions which are not about anything except themselves? That is not the impression that one gets from reading Wittgenstein or Ryle. On the other hand, Ayer's claim that "the propositions of philosophy are not factual, but linguistic in character " (Ayer,1936, p.35) opens the door to the usual questions. If language is not factual, how can we tell whether or not we are merely spinning our wheels by analyzing language as a method of doing philosophy? How can we tell the "factual" from the "linguistic"? How can we tell that the language we use in our own analysis of language is not "merely linguistic"? These issues also come up in literary criticism. If language is not about reality, it follows that literature must not be either. Thus we have the concept of "literary autonomy." J. Hillis Miller (1971 ), for example, puts it this way:

If meaning in language rises not f rom the reference
of signs to some thing outside words but, from differential
relations among the words themselves, if "referent "
and "meaning"must always be distinguished, then the
notion of literary text which is validated by its one-to-one
correspondence to some social, historical, or psychological
reality can no longer be taken for granted(p.85).

It appears that this ideology has, to some extent become a self-fulfilling prophesy. Much of modern literature is the result of a conscious attempt to avoid the description of any external reality. The extreme form is modern poetry which consists of the juxtaposition of words in a manner which defies being read in anything like the normal manner. The reader looks in vain for any meaning at all. Modern critics, however, are not content to restrict this "formalist" interpretation to works which are explicit in their formalism. What must count as the reductio-ad-absurdum in this vein are such statements as Miller's claim that Charles Dickens' *Sketches by Boz* only seems to be an example of straight-forward realism. In truth, says Miller, it was something of a Trojan Horse to illustrate the fallacies of realism. The work calls attention to its own fictionality. He tells us that "This movement lies, between a focus upon the literal referents and a focus upon their fictional quality challenges the authenticity of what is represented while what is represented in its turn undermines the apparent solidity of the Sketches as an innocent act of representational mirroring" (pp. 115-116). And later we find that the narrator "gives the reader the information he needs to free himself from a realistic interpretation of the *Sketches*"(p.119).

Analogous to this New Criticism in literature is the approach to art criticism championed by Clement Greenberg. In modern art, according to Greenberg (1961,p.6), "Content is to be dissolved so completely into form that the work of art or literature cannot be reduced in whole or in part to anything not itself." Thus, "avant-garde culture is the imitation of imitating" (p.8). Also like Miller, Greenberg was not above reinterpreting realistic paintings into formalist terms. For example, he suggests that critics of the future may consider the illusion of depth and volume (in the Old Masters) to have been aesthetically valuable primarily because it enabled and encouraged the artist to organize such infinite subtleties of light and dark, of translucence and transparence, into effectively pictorial entities (pp.137-138). The radical skepticism which is born of a misapplication of the scientific endeavor comes full circle in modern philosophy of science. The philosopher of science's equivalent to Locke's distinction between primary and secondary qualities is the distinction between fact and theory or between observational statements and theoretical statements. The former are real. The latter are inferred from the former. The logical positivists tried to ground science in a solid, proven set of observables which were not themselves dependent upon any particular theory. However, the realization that the positivists' conception of observation was not powerful enough to account for the facts of science opened the door to a skepticism from which the philosophy of science has not yet recovered. As Lakatos (1970) describes the predicament:

The demarcation between the soft, unproven "theories" and the hard, proven "empirical basis" is nonexistent: all propositions of science are theoretical and, incurably, fallible (p.100).

Thus, the belief that science is uniquely capable of getting at the truth engenders a skepticism concerning non-scientific truth which is inevitably turned upon science itself. The extreme form is so-called "conventionalism," the belief that the

facts of science are merely conventions which communities of
scientists agree to treat as real.
In sum, dualism and its offspring, idealism and skepticism,
are very much a part of modern life. We have argued that this is
the result of giving favored ontological status to the elements
of physics. Of course, the picture is not quite that simple.
As P.G.Wodehouse was won't to have his characters say, "There
are wheels within wheels." The various fields have been influenced
by each other and by our comnmon philosophical heritage
as well as directly by physics. And the influence is typically
from a complex set of conceptions and misconceptions about the
nature of the physical sciences. Likewise, the demonstration
that the 'physical' world is not important for a given phenomenon
often goes hand in hand with a desire to turn the tables
and reduce everything to that phenomenon. So the literary
critic would like to demonstrate that all is 'fictional', the biologist that
all is cell activity, and the linguist that all is
language. Despite these complications, however, we have intended
to demonstrate that a certain attitude toward physics is (and
has been for centuries) the root cause of radical skepticism.
The problems with radical skepticism are manifold, as
philosophers such as G. E. Moore , J . L. Austin, and Gilbert
Ryle have shown in such detail (e.g., Moore, 1922, Ryle, 1949;
Austin, 1962). The obvious problem, of course, is that it is
self-defeating. If we cannot know anything, how can we know that
we cannot know anything? There is also the problem of infinite
regress. A statement like "All X is about X" leads either to
infinite regress or becomes a meaningless tautology, or both.
Of course, this infinite regress can be elevated to the status
of a general principle such as "self-referentiality," but that
does not make it any less an infinite regress. Another problem
is that the doctrine of "autonomy" has the effect of trivializing
what it attempts to illuminate. If art is only about
itself, why should nonartists bother with it? Why should we as
a culture support it? We could go on, but this paper is not
really addressed to the committed skeptic, who would not, by
his own criteria, accept our (or anyone else's ) arguments as
true. Therefore, let us go on to a discussion of three episodes
in science which epitomize the eternal dialectic between the
Euclidean and Non-Euclidean approaches to the universe.

### 3. Euclidean vs. Non-Euclidean

Theories of light and sound are among the most basic and important of the sciences.

This is easy to understand given the central role of seeing and hearing in human experience. Optics and harmony have held a central place in the mathematical sciences since their beginnings in ancient Greece. The early Pythagoreans tried to account for the pitch of a note by means of simple ratios between small whole numbers. Their enthusiasm for this approach to harmony was expressed in their slogan "all is number." This approach or vision is one of the main characteristics of Western mathematical science and has often been very successful. But right from the beginning it had its opponents. In the 4th century B.C. Aristoxenus opposed the Pythagorean theory of harmony. Van Jan has summarized Aristoxenus' teachings as follows (the material is quoted from Szabo, 1978, pp.11 2-113):

In his harmonics he did not inquire into the origin of a tone, nor did he ask whether it was a number or a speed. The ear need only listen unaffectedly to the range of tones; it will tell us with certainty which tones harmonize with each other. His system is based on the fourth and the fifth which are easily perceived consonances, and without asking which numerical ratios underlay them, he was able to determine from them the whole and half tones, etc.

As a matter of fact, Aristoxenus' own words also attest to his anti-Pythagorean tendencies:

We are attempting to draw conclusions which are in agreement with the data, as opposed to the theorists who proceded us. Some of them introduced completely foreign viewpoints into the subject and dismissed sensory experience as imprecise; hence they made up intelligible causes and stated that there were certain ratios between numbers and speeds on which the pitch of a note depended. These were all speculations which are completely foreign to the subject and absolutely contrary to appearances. Others renounced reasons and arguments completely and proclaimed their assertions as though they were oracular sayings; they likewise paid insufficient attention to the data.

Several centuries later Goethe responded to Newtonian optics in a manner reminiscent of Aristoxenus' response to Pythagorean harmony. Mathematical optics had made great progress in the 17th century, culminating in Newton's theory of color expounded in his book *Optics*. We first quote from Newton's (1730) *Optics* so that the reader can see what Goethe was responding against.

The homogeneal light and rays which appearred, or
rather make objects appear so, I call rubrific or
red-making; those which make objects appear yellow,
green, blue, and violet, I call yellow-making, greenmaking,
blue-making, violet-making, and so of the
rest. And if at any time I speak of light and rays
as coloured or enbued with colours, I would be understood
to speak not philosophically and properly, but
grossly, and according to such conceptions as vulgar
people in seeing all these experiments would be apt
to f rame. For the rays, to speak properly, are not
coloured. In them there is nothing else than a certain
power and disposition to stir up a sensation of this or
that colour. For as sound in a bell or musical string,
or other sounding body, is nothing but a trembling
motion, and in the air nothing but that motion propagated
from the object, and in the sensorium 'tis
a sense of that motion under the form of sound; so
colours in the object are no thing but a disposition
to reflect this or that sort of rays more copiously
than the rest; in the rays they are nothing but their
dispositions to propagate this or that motion in the
sensorium, and in the sensorium they are sensations
of those motions under forms of colours (p.428).
All the colours in the universe which are made by
light, and depend not on the power of imagination,
are either the colours of homogeneal lights, or compounded
of these, and that either accurately or very
nearly, according to the rule of the foregoing problem(p.442).

The reader should note the "nothing buts," the primary-secondary
distinctions, and the disdain for the ordinary person's
viewpoint. (There is, of course, also much of great
worth in Newton's *Optics* which even Goethe wouldn't have
denied). Now for Goethe's reaction (the quotes are taken from
(Goethe, 1971):

The dyer, a technician, will welcome our efforts.
Those who have considered on the phenomena of dyeing
are particularly unsatisfied with the current theory.
They were the first to become aware of the shortcomings
of Newtonian theory. It makes a great difference
how one attacks knowledge, a science, through
which door one enters. The practitioner, the manufacturer,
who is daily and forcefully impressed by these phenomena,
who experiences utility or harm by applying his convictions,

who is not indifferent to losses of money and time but wants to go on
to other accomplishments -- it is he who discovers much more
rapidly than a scholar the fallacy of a theory, he
who in the end must regard theory as honest currency.
He cannot operate like the mathematician, whose formula
is infallible even though the experience on which
it is based is wrong. He will also approach the color
theory from the viewpoint of painting, the aesthetic
coloration of surfaces, and will hope to accomplish a
most thankworthy task for the artist. We have endeavored
to define in the sixth part the effects of color as
addressed at once to the eye and mind, with a view
to making them more available for the purposes of art.

Although much in' this portion, and indeed throughout,
is only a sketch, it should be remembered that
all theoretics can in all strictness only point out
leading principles, under which guidance, practice
may proceed with vigor and be enabled to attain legitimate
results (p.77).

It is blasphemy to say that there is optical deception
(p.38).

The creative artist could gain but little advantage
from, a theory whereby the optical scientist by his
negative efforts merely explains overall occurring
phenomena. Even though he admires the various colors
of the prism along with other observers and has
invented the harmony of these colors, it still remains
a mystery to him just how he should achieve his objective,
based on certain color relationships that he has
created and organized. -The harmony of a painting depends
to a large degree on light and shade. But the
relationship of color to light and shade is not so
easily discerned. Yet every painter soon discovers that
the mere combination of both harmonies can fully complete
his painting. It is not enough to mix color with
black or brown in order to make a darker shade. Many
attempts to innately gifted eyes, the exercise of
sensitivities, and the tradition and example of great
masters finally bring artists to a high plateau of excellence.
Yet these artists could scarcely communicate the rules upon
which they operate. One can convince oneself while viewing
a great collection of paintings that nearly each master had a different
way of handling color (p.18).

It is clear that Goethe was concerned with our entire experience of color and saw almost no advantages -- and in fact many disadvantages -- in Newton's reductionistic, "nothing but" approach. Recently, the psychologist J. J. Gibson has taken the same attitude towards the problem of perception. Gibson is quite explicit in his rejection of traditional physics as part of a theory of visual perception.

The size-levels of the world emphasized by modern physics, the atomic and the cosmic, are inappropriate for the psychologist. We are concerned here with things at the ecological level, with the habitat of animals and men, because we all behave with respect to things we can look at and feel, or smell and taste, and events one can listen to (Gibson, 1979, p.9).

Gibson advocates what he calls an ecological physics: redescription of the physical world relevant to perceiving organisms. An interesting thing about Gibson's position is that, while not denying the reality of the physicist's description, he considers this ecological description to be just as 'real,' just as 'physical', and in fact more relevant for the particular purposes of the psychologist. Gibson's central theme might be paraphrased "What you see depends upon what you have to do." Different organisms perceive different properties of the world, depending upon their particular needs. Thus, each organism, in effect, lives in a different environment, or what Gibson calls an "affordance structure." As Gibson puts it, "The affordances of the environment are what it offers the animal, what it provides or furnishes, either for good or ill" (p.127). Thus, high frequency sound is part of the bat's world but not of the human's (unless he is a sonar operator), and well defined color is part of the human's world but not the bat's. But it is important to emphasize that this relativity does not in any way imply that the environment is "subjective" or exists only in the "organism's head." Gibson calls his position "direct realism." In many ways, it is an articulation of the sort of position Gilbert Ryle argued for. It follows from simply taking as real what people say they see as real. Gibson's position holds out the possibility of avoiding skepticism while still allowing for fundamental incompatibilities between viewpoints. In the next section we develop a position related to Gibson's in the attempt to solve the problems we described in the previous sections.

## 4. Perspective Realism

One way of describing the differences between the various sciences is to say that they have different conceptions of the line between primary and secondary qualities. In this sense, Einstein's general relativity is, perhaps, the most extreme view in one direction, in that the geometry of space-time is all that is given primary status. Everything else becomes an epiphenomenon. On the other hand, Gibson's ecological psychology may be the most extreme view in the other direction. Gibson's program might be described as an attempt to eliminate the notion of secondary qualities altogether, in the same sense that Kant, for example, tried to eliminate the primary qualities. This liberalism of Gibson's recommends his direct realism for use as a metaviewpoint which might encompass all the others. In this section we will pursue this idea.
We have seen that there is a plurality of viewpoints, and that the attempt to deny this plurality (by giving favored ontological status to one viewpoint) leads to solipsism. Our goal here is the development of a general account which will neither deny the reality of the world as described by the various sciences, nor deny the reality of the world of common sense. Our basic theme is that any distinction between primary and secondary qualities is only valid relative to some specific viewpoint. That is, each way of looking at things defines its own primary-secondary distinction and thus its own ontology.
The relationship between scientific observation and scientific theory is formally similar to the traditional distinction between sensation and perception. As we have seen, the traditional notion of sensation was the way that physics was allowed into psychology. It served a felt need for a mental description which was commensurate with the physicist's world and thus the same for all knowers. The particular description which evolved had the added appeal of being apparently compatible with receptor physiology. Modern psychologists argue about the details of sensation, just as the British Empiricists had before them -- whether we "really" see patches of color or merely points of light - - however, before Gibson there was little doubt of the need for the concept of sensation. According to Gibson, however, the concept of sensation is an unfortunate confusion between the scientist and the ordinary perceiver. There is no evidence whatsoever for a primitive experience of meaningless points of light or colored shapes f rom which we infer our perceptions. By eliminating the concept of sensation altogether, Gibson is able to reject this indirect realism which leads to skepticism. Gibson actually does what philosophers had been claiming to do from the time of Descartes, he accepts as real the comman man's description of what he experiences. It follows that different experiences of the "same thing" are real differences and not merely different inferences from the same sensations. To reiterate, for Gibson, what is experienced is the affordance structure of the environment, namely the implications of the experiencer's potential actions. Experiences differ because potential actions (i.e., goals, purposes) differ. Another important idea of Gibson's is that experience itself evolves in connection with action, so it makes sense to speak of organisms learning to experience the world in ways appropriate to their purposes.

Although Gibson's work was concentrated on what has traditionally been called perception, his approach constitutes a general theory of mind. It is a theory of mind which treats all mental activity as a direct apprehension of the world. So just as perception is not built upon sensation, thinking and memory are not built upon perception. Rather, these "higher mental processes" are the apprehension of different properties, of the world, e.g., those that we refer to as past. Thus there is no distinction between seeing and seeing as. There is just a direct apprehension. All seeing is already a seeing as. Of course, it goes without saying that this approach requires a great deal of flexibility concerning what exists. Indeed, from this point of view, one might say that what is seen is what exists. Now let us turn to a consideration of scientific observation. In the same way that the traditional notion of sensation constrained models of perception, the traditional notion of observation constrains models of science. Carnap's program was the classic example of the attempt to reduce science to a preconceived idea of observation. In a sense it was an attempt to reduce science to sensation in that his simple observation statements were very similar to the age-old description of sensations. But, as with all grand reductionist programs, Carnap's left a huge gap which had to be filled with "induction" or some such mechanism to get from his simple observation statements to higher-level theoretical statements. For example, from Carnap's point of view, atoms cannot be seen, they must somehow be inferred from simple observations. Another problem with Carnap's approach is that even if possible in principle, the actual working out of such a reduction is intractable in practice. This is where Gibson comes in. Suppose we say that scientists can see atoms, that physicists use their instruments as tools to see atoms in the same way that astronomers use telescopes to see stars. There is a tendency to say that all you really see is the face of a dial or some equations. However, this objection assumes that there is one level of analysis which can without confusion be called observation. An analogy will help to show what is wrong with this preconceived notion of observation. When listening to a radio, what do we hear? We can identify as the direct experience any number of steps in the causal sequence. There are vibrations of the speaker, radio waves coming through the air, electrical impulses in the microphone, vocal cord vibrations of the disk jockey, etc. Where we plug in to this causal nexus depends upon the kind of account we want to give. Under most circumstances we would say that we hear what the disk jockey is saying. The other phenomena are merely the mechanism by which we accomplish this. It would make little sense to say that all we really hear are radio waves or speaker vibrations. Although such a statement would be true in a certain sense, it would be quite misleading for most purposes. Likewise, it is quite misleading to say that all a scientist sees are the positions of his dials. On the contrary, scientific training is designed to insure that scientists see a great deal more than the positions of dials. The fact that the nonscientist only sees dials only means that he has not learned to see as a scientist. It is not different from the inability of a non-chess player to see a sophisticated defense rather than just pieces being moved across the board.

Therefore, let us say that different sciences have truly different viewpoints because they have different purposes. There is a general viewpoint for a given science (i.e., a metatheory) which is presupposed in the collection of data and which is connected with the goals and purposes of the science. The collection of data then becomes a process of looking at the world in a particular way.

It follows that the physicist's world is one way of looking at things, but **not** the way of looking at things. The physicist's viewpoint is relevant (indeed extremely powerful) for certain types of manipulations, predictions, etc., but useless for others. The same could be said of the world as described by any other science.

Another analogy might be useful. Imagine a group of objects which vary subtly both in size and weight. Suppose some one is asked to classify the objects by size and given a very sensitive pair of calipers. Another person is given an extremely accurate scale and asked to classify the objects by weight. Neither classification is arbitrary, but the two are totally incommensurable. Weight simply does not exist for one person, size does not exist for the other.

We could make the situation more interesting by making their respective instruments appear to be identical, complete with the same names for the result of the measurements. Imagine the arguments which would ensue when the two got together to discuss the objects in question. The problems might even be aggravated by the fact that weight and size are often correlated.

Now let us generalize this example. The basic idea is that the world contains a great deal of structure which can, be "sliced up" in many different ways, depending upon one's needs. Thus the world from any particular point of view has a structure which contains some but not all of the total possible structure. To formalize this a bit we could say that there is a many-one mapping relation between the external world and the particular world specified by a given viewpoint . This differs from the more traditional position which relies upon the one-many mapping relation to account for differences in viewpoints. The traditional position is that there is only one viewpoint which is real (and which all perceivers have access to), but that various things are inferred from or added to this viewpoint depending upon the perceiver's characteristics. The advantage of our formulation is that it avoids the pessimistic conclusion that the individual worlds defined by particular viewpoints are at most only inspired by the world. To approach this issue from another angle, our formulation avoids the bizarre conclusion that each individual viewpoint is actually more complex than the world. To continue our scheme, these products of many-one mapping relations we call "homomorphisms" (as opposed to isomorphisms), following Ashby (1956). Now the different homomorphisms of an original set can be arranged into the mathematical structure called a lattice. An example of the lattice formed of an original set and its homomorphisms is shown in figure 2.

2.

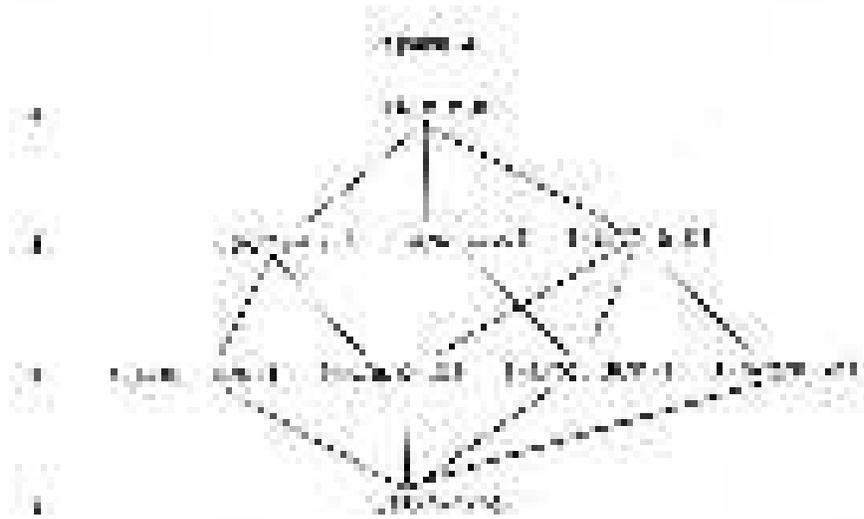

The figure represents a set containing four elements which generates
eight homomorphic sets (there are six other possible homomorphisms
which we have left out for simplicity). The parentheses
indicate that the elements within them are not distinguished.
The vertical lines indicate which sets are homomorphisms of
which other sets at the higher levels. The numbers to the left
signify the number of separate elements.

So the original set contains four elements; the first generation of homomorphisms
contains three elements, and so on.
Notice that although all of the sets are veridical simplifications
of the original, there are very basic incompatibilities between many
of them. Another important thing to notice is that the degree of overlap
between sets and the relative number of elements are orthogonal dimensions.
Two sets which contain an equal number of elements may or may not be similar.
A set which is simpler than another (i.e., has fewer elements)
may be a homomorphism of it, they may be totally unrelated, or
there may be partial overlap. Thus, different viewpoints may be
related (or unrelated ) in a number of different ways, and yet all be "true"
in the sense that they are all accurate simplifications of the world.

Now let us turn to a pair of basic issues which have traditionally
been troublesome for philosophers of science: the
problem of error and the problem of progress. Both are more
specific forms of the incompatibility which we have dealt with so far in a general way. The key to the problem of error is that it is viewpoint specific. A more formal way of putting this is to say that the truth value of any statement cannot be determined until the viewpoint is specified. One type of error is the simple failure to make a distinction which is relevant. Another type is to adopt a viewpoint which is inappropriate to the task at hand. In general an error is something which hampers the basic goals to which the viewpoint is specific. For example, a driver who fails to see a truck which pulls out in front of him would be said to have made an error despite the fact that he may have looked at that region of space-time in any numbe r of ways which would have been appropriate in other circumstances. This example illustrates an important point - -a truck is not really a collection of atoms. A t ruck is a truck. It has implications which are not contained within the notion of "collection of atoms." Collections of atoms do not pullout in f ront of people, trucks do.

The issue of scientific progress is complicated by the admission that there are fundamentally different goals and thus fundamentally different viewpoints. The traditional notion of progress is best exemplified by the case in which the goals remain relatively constant and more and more distinctions are made. This would be equivalent to "moving up the lattice." On the other hand, goals change as well, that there can be radical transformations of viewpoint which cannot easily be called progress. In truth, both of these processes are probably taking place constantly, so that there is continuous refinement of the viewpoint relative to particular goals and continuous reassessment of goals. It should be clear f rom what we have said so far that communication is only possible within a viewpoint, not between them. The only way to foster communication between two viewpoints is to adopt a third which encompasses the first two. Likewise, there can be no commensurability of viewpoints until there is agreement about goals. This is a particular problem when those who hold two viewpoints do not realize that their goals are different.

In the next two sections we take another look at the "hard" sciences. We will see that even in the heart of the hard sciences there is a multiplicity of viewpoints. This multiplicity is not some thing to be bemoaned, but is instead the natural result of a plurality of goals and computational capabilities.

**5. The Exact Sciences and Non-Euclidean Logic**

As we have seen, the question of what "really" exists pervades
the sciences and human thought in general. The belief
that the infinite does not really exist goes back at least to
Aristotle. Parrnenides even questioned the reality of plurality
and change. (Einstein's vision has much in common with
Parmenides). Towards the end of the nineteenth century an
acrimonious exchange took place between Kronecker and Cantor
regarding the reality of the actual (as opposed to potential)
infinite. Kronecker claimed that only the finite integers really
exist and all else is merely the work of man. Cantor countered
that the essence of mathematics was its freedom and that he had
attained a larger vision than Kronecker had who could not see the
infinite. Most mathematicians have followed Cantor and found his
paradise a more beautiful and alluring universe.
Hilbert accepted Kronecker's viewpoint for his metalanguage,
but tried to recapture Cantor's paradise in a formal
language. Hilbert was a formal pluralist in feeling that each
mathematical discipline was entitled to its own formalization.
Russell was a logical monist and felt that all of mathematics
should be constructed within a single formal system. He put a
great deal of labor into his program and looked askance at
Hilbert. He felt that Hilbert's approach had all the advantages
of theft over honest toil. What he did not realize was that in
intellectual affairs, as in economic affairs, great fortunes
are rarely ever accumulated through honest toil. What is needed
is the intellectual leap. Russel's program led to much interesting
mathematics, but even if in principle it could be
carried out, in practice the result would be computationally
intractable. One would be translating simple, clear ideas into
the fog of *Principia Mathematica*. Russell's program has as much
relevance to complex analysis as von Neumann's game theory has
to chess. The understanding and appreciation of mathematics has
very little to do with formal logic. For example, the following
footnote occurs at the beginning of Wall's (1970) book *Surgery
On Compact Manifolds*.

Recent results of Kirby, Siebenmann and Lees have now
(1966) provided such a technique. All our methods now
extend to the topological case, with only trivial alteration.
See (K8),(K9), (L10).

All the experts could see the truth of this footnote. But
this seeing is not explained by modus ponens. In his beautiful
book *Proofs and Refutations*, Lakatos (1976) has shown that the
mathematical process itself is dialectical and not Euclidean.
At all times our ideas are formally inconsistent. But inconsistency,

while still recognized as a pathology, is no longer
seen to be a fatal disease. If we come across a contradiction,
we localize it, isolate it, and try to cure it. But we have to
get over our neurotic phobias concerning this disease and recognize
it as inseparable from life itself.
Hilbert's program collapsed with the startling work of.
Godel. Mathematical logic and the study of formal systems have
become a branch of mathematics instead of its foundation.
Moreover, Cantor's paradise has been raised into the metalanguage
in order to prove deep theorems concerning formal systems
as well as to provide a semantics for such systems. A. Robinson
even defined non-standard formal systems which contain infinite
formulas. One thus has a large plurality of different approaches
to mathematics. Most mathematicians live in Cantor's paradise
in spite of Russell's paradox; they simply learn to avoid making
certain moves which have been shown to lead to contradictions.

The situation is similar in physics also. As soon as the
phenomena under discussion become sufficiently complex, one must
depart from Euclidean strategies and adopt a non-Euclidean
approach. This is very clearly stated by Blandford and Thorne
(1979, p. 454-460) in the context of black hole astro-physics.

The fundamental theory of black holes, as laid out in
chapters 6 and 7, is well posed, elegant, clean and self contained.
It follows inexorably and clearly f rom the
fundamental laws of physics. The theory of black holes
in an astrophysical environment is completely the
opposite. Because it deals with the physics of matter
in bulk-matter orbiting and accreting onto a hole- -
it is subject to all the dirty, complex uncertainties
of the modern theory of the behavior of bulk matter.
If thunderstorms and tornados on Earth have eluded
accurate theoretical modeling, how can one expect to
predict even qualitatively their analogues in the
turbulent, magnetized plasmas that accrete onto a
black hole in a close binary system? One cannot.
The best that can be hoped for is to develop the crudest
of models as to the gross behavior of matter in the
vicinities of holes. Fortunately, the resulting models have
some modest hope of resembling reality. This is because
the relative importance of physical processes near a hole can be
characterized by dimensionless ratios that usually
turn out to be very large, and consequently, the
gross behavior of matter near a hole is dominated by
a small number of processes. The task of the model

builder is to identify the dominant processes in his
given situation, and to construct approximate equations,
describing their macroscopic effects. Historically,
to identify the dominant processes has not been easy.
This is because a vast number of possible processes must
be considered and the model builder often, out of iqnorance,
overlooks an important one.
Thus it is that research on black hole astrophysics
involves large bodies of physical theory. Within each
body of theory one must have at one's fingertips
approximate formulae that characterize a long list of
possibly relevant processes. The necessary bodies of
theory include general relativity, the physics of
equilibrium and non-equilibrium plasmas, the physics of
radiative processes, and the physics of stellar dynamical
systems.
Research in black hole astrophysics also requires, a good
knowledge of the phenomenology of modern astronomy – the
observed properties of stars, the main features of
their evolution, the structure of the Galaxy, and the
observed physical conditions in interstellarspace.
The above example illustrates the following features
of research in black hole astrophysics:
(i) It involves an iteration back and forth
between the equations of the macroscopic model and the
microscopic physics which underlies those equations.
One iterates until one obtains self-consistency.
(ii) One must search carefully, at each iterative
stage, for overlooked processes that might be so important
as to invalidate the model (anchoring to a homogeneous
interstellar magnetic field in the above example).
(iii) One frequently encounters a 'branch point'
where the model will take on two very different forms
depending on what one assumes for the environment around
the hole (homogeneous magnetic field versus tangled field
in the above example), and where both branches might
well occur in the real universe. This leads to a
plethora of possible models, each corresponding to a
different black hole environment and/or range of black
hole masses.

Probably the clearest case study of non-Euclidean logic
occurs in the paper *Problem Solving About Electrical Circuits*
by Stallman and Sussman (1979, p.33-39). We quote from their
introduction.

A major problem confronting builders of automatic
problem-solving systems is that of the combinatorial
explosion of search-spaces. One way to attack
this problem is to build systems that effectively use
the results of failures to reduce the search space
that learn from their exploration of blind alleys.
Another way is to represent the problems and their
solutions in such a way that combinatorial searches
are self limiting.
A second major problem is the difficulty of debugging
programs containing large amounts of knowledge.
The complexity of the interactions between the "chunks"
of knowledge makes it difficult to ascertain what is to
blame when a bug manifests itself. One approach to
this problem is to build systems which remember and
explain their reasoning. Such programs are more convincing
when right, and easier to debug when wrong.
ARS is an expert problem-solving system in which
problem solving rules are represented as demons with
multiple patterns of invocation monitoring and an associative
data base. ARS performs all deductions in an
antecedent manner, threading the deduced facts with
justifications which mention the antecedent facts used
and the rule of inference appl ied. These justifications
are employed by ARS to determine the currently active
data-base context for reasoning in hypothetical situations.
Justifications are also used in the analysis
of blind alleys to extract information which will limit
future search . . .
ARS supplies dependency - directed backtracking, a
scheme which limits the search as follows: The
system notes a contradiction when it attempts to solve
an impossible algebraic relationship F, or when it
discovers that a transistor's operating point is
not within the possible range of its assumed region.
The antecedents of the contradictory facts are scanned
to find which nonlinear device state guesses (more
generally, the backtrackable choicepoints) are relevant.
ARS never tries that combination of guesses again. A
short list of relevant choicepoints eliminates from
consideration a large number of combinations of answers
to all the other (irrelevant) choices. This is how
the justifications (or dependency records ) are used
to extract and retain more information f rom each
contradiction than a chronological backtracking

system. A chronological backtracking system would
often have to try many more combinations, each time
wasting much labor rediscovering the original contradiction.

Note, that this is a case where a Euclidean approach does
exist but is computationally intractable. We thus have a clear
example of a non-Euclidean approach being used for good reasons
and not just because of sloppiness. We thus gain a better
appreciation of the gulfs that separate the mathematician,
the physicist, and the engineer. Their programs are different,
their aesthetics are different, and even their "logics" are
different. At this point, the question of whether every non-Euclidean
logic can be embedded in a Euclidean logic can only
be answered dogmatically. Those in artificial intelliqence
research assume a positive answer, while hermeneuticists assume
a negative answer. However, even if you believe in an affirmative
answer, a very simple non-Euclidean logic may only be
embeddable in an immensely complex Euclidean logic, thus making,
its embeddability irrelevant from a practical point of view
(recall von Neumann's analysis of chess).

## 6. Biology and Modern Physics

Modern physics pervades nearly all aspects of modern biology.
The striking difference between the explanations given by
Aristotle and by modern biologists is partially explained by
the fact that cells, molecules, atoms, electrons and photons
were not part of Aristotle's universe. Modern biology teaches
us that we breathe in order that oxygen can be supplied to the
cells of the body, so that they can metabolize food to obtain
energy needed for the other activities of life. This helps us
understand the structure and purpose of the lungs, red blood
cells, etc. Much of this understanding is at the level of
molecular structure (for instance, the structure of hemoglobin).
We explain the workings of green plants by saying that they are
green because of the presence of molecu l e s of chlorophyll which
are used to capture photons from the sun, thus providing the
plant with its basic energy supply. The resemblance of children
to their parents is explained by the DNA-sequence in their
chromosomes. Sight is explained by appeal to photons captured
by the rods and cones in the retina. One could go on and on.
Considering modern physics' pervasive role in modern
biology, it is quite interesting to note how little most biologists
know of the foundations of modern physics. Their ideas of photons, atoms, and molecules resemble the incoherent pictures formed by Einstein and Bohr during the period 1900-1925. Even more interesting is the fact (?) that their ignorance of the viewpoints of modern quantum mechanics does not seem to hinder their work at all. While they think of photons and electrons "classically," the success of their work is due to the fact that they only actually use atomic language in the way Ostwald recommended at the end of the nineteenth century. This can be easily seen in Watson's book *The Molecular Biology of The Gene*. Let us reconsider our previous biological examples using correct quantum mechanical language. First of all, photons are not objects which could come from the sun. Instead, one assumes the existence of certain detectors called photon detectors which give positive readings under certain conditions such as being exposed to the sun light. Next, one discovers that plants grow well not only in sun light but also under many other conditions. A basic invariant of all these conditions is that under them our photon detector would yield positive results. Again, molecules aren't objects either. So plants aren't "made up" of molecules such as chlorophyll. Instead, another basic invariant of those conditions under which plants grow successfully and appear green is that a chlorophyll detector would yield a positive result. Children don't resemble their parents because of DNA sequences, but, instead, parents and their children's DNA-sequences are highly correlated and each is somewhat correlated to phenotypes. This is not the type of language usually used by biologists. But biologists like Watson aren't wholly to blame for their misuse of physical language. Watson gets his physics from Linus Pauling. In his book, *The Nature of the Chemical Bond*, Pauling constantly mixes classical ontology with modern quantum mechanics. Consider for example the following quote (Pauling, 1960, p.19).

The electron distribution function for molecule-ion is shown in Figure 1-5. It is seen that the electron remains for most of the time in the small region just between the nuclei, only rarely getting on the far side of one of them; and we may feel that the presence of the electron between the two nuclei, where it can draw them together, provides some explanation of the stability of the bond.

This picture resembles more the hidden variables views of Bohm and Nelson than it does the viewpoints of Bohr and Heisenberg. But the master knew better as is shown by the following

quotes (Paulinq, 1960, pp. 217- 220).

It is true that chemists, after long experience in the use of classical structure theory, have come to talk about, and probably to think about, the carbon-carbon double bond and other structural units of the theory as though they were real. Reflection leads us to recognize, however, that they are not real, but a rhetorical constructs in the same way as the individual Kekule structures for benzene. It is not possible to isolate a carbon-carbon bond and to subject it to experimental investigation. There is, indeed, no rigorous definition of the carbon-carbon double bond. We cannot accept, as a rigorous definition, the statement that the carbon-carbon double bond i s a bond between two carbon atoms that involves four electrons, because there is no experimental method of determining precisely the number of electrons that are involved in the interaction of two carbon atoms in a molecule, and, of course, this interaction has rigorously to be considered as being dependent on the nature of the entire molecule. I feel that the greatest advantage of the theory of resonance, as compared with other ways (such as the molecular-orbital method) of discussing the structure is that it makes use of structural elements with which the chemist is familiar. The theory should not be assessed as inadequate because of its occasional unskillful application. It becomes more and more powerful, just as does classical structure theory, as the chemist develops a better and better chemical intuition about it.
The theory of resonance should not be identified with the valence-bond method of making approximate quantum-mechanical calculations of molecular wave functions and properties. The theory of resonance is essentially a chemical theory (an empirical theory, obtained largely by induction from the results of chemical experiments). Classical structure theory was developed purely from chemical facts, without any help from physics. The theory of resonance was also well on its way toward formulation before quantum mechanics was discovered.
The theory of  resonance in chemistry is an essentially qualitative theory, which, like the classical structure theory, depends for its successful application largely upon a chemical feeling that it developed through

practice. We may believe the theoretical physicist who
tells us that all the properties of substances should be
calculable by known methods-- the solution of the
Schrodinger equation. In fact, however, we have seen
that during the 35 years since the Schrodinger equation
was discovered only a few accurate nonempirical quantummechanical
calculations of the properties of substances
in which the chemist is interested have been made. The
chemist must still rely upon experiment for most of his
information about the properties of substances.
Experience has shown that he can be immensely helped by
the use of the simple chemical structure theory. The
theory of resonance is a part of the chemical structure
theory, which has an essentially empirical ( inductive )
basis; it is not just a branch of quantum mechanics.

Thus, no matter what the future may bring, chemistry is at
present an independent science only slightly dependent upon
physics and much dependent upon chemical intuition. The situation
is similar for biology. Watson goes out of his way to
insist that biology requires no "natural laws" not already found
by chemistry and physics. He is opposing vitalists such as
Bergson and the expectations of physicists such as Bohr and
Schrodinger. Bohr hypothesizes that biology would yield a
biological complementarity principle where the description of
a living organism would be complementary to a description of its
molecular structure. Such a position may yet come to pass when
biology becomes sufficiently precise to become conscious of its
"complementarities."

**7. Natural Supernaturalism**

While there is a plurality of viewpoints within the sciences,
the gap between the sciences and the humanities is much larger
than that within the sciences. All the sciences attempt to
take a value neutral, I-it, approach to their respective
subject matters. This is as true in anthropology as it is in
physics. An anthropologist qua scientist attempts to describe
man and his culture in unemotional language similar to the
astronomer's description of the planets and their motion. But
the use of such language is often found very objectionable by
non-scientists. The astronomer has succeeded in demoting the
planets from gods to merely large chunks of rock. Humanists
are often afraid that science will have a similar "success"

with man. Plato railed at the atomists for presenting a vision of the universe stripped of its most important aspects: value, honor, justice. The debate continues today with such thinkers as B. F. Skinner and E. O. Wilson. The effects of taking the scientific vision too seriously are clearly seen in Pascal's fleeing it in horror and more recently in the fatalistic tone of Loren Eisley's autobiography. The scientific vision inevitably leads to viewing life as " .. . a tale/ told by an idiot full of sound and fury/ signifying nothing." At the conclusion of his book *The First Three Minutes*, Steven Weinberg (1977) unconsciously recognizes the limitations of the vision he has presented when he suggests that the acceptance of this vision is one of the few things that raises the human predicament from a farce to a tragedy. But viewing the human predicament as either a farce or a tragedy is a point of view which does not fit into the framework of the big bang model. The universe Weinberg presents is simply not the total universe, but merely the (or better yet, a ) physical universe.

At the beginning of the nineteenth century, romantics such as Wordsworth staged a counterattack on the scientific world view. The old Christian worldview of the *Bible*, Dante and Milton was no longer believable after the scientific revolutions of the seventeenth and eighteenth centuries. But, on the other hand, it was also clear to the romantics that Newton was not enough. Unfortunately, in their counterattack they often went off the deep end. Man was no longer at the center of the universe but was instead its creator. Fichte's ego resembles Plotinus' God with the universe an emanation thereof. This is certainly more than ample compensation for our loss at the hands of Copernicus. But is the Milky Way really something we can ask credit for? To reiterate a point, the problem with many of the romantics was that they conceded too much to the physicists. They gave the physicists the real world and were willing to settle for the mind of man. But is beauty any less in a sunset than its color? What is needed is a vision of the world which includes beauty and justice on the same ontological level with color and mass. Lack of consensus with regard to these values simply means that there will be a plurality of such visions - -as Mao said (but didn't mean) "let a thousand flowers bloom." Insistence on consensus merely results in impoverished world views. In the next section we consider some of the richest worldviews, namely, those including the supernatural.

**8. Supernaturalism Revisited**

When Napoleon asked LaPlace where God was in his universe, LaPlace responded that he had no need of that hypothesis. Since the publication of Darwin's evolutionary theories, all the force has gone out of the argument from design. Aristotle's argument of the first cause or unmoved mover has always been unmoving. But, for many people, God is not an hypothesis, but a fact, a phenomenon, an experience. For these people LaPlace and Darwin and secular humanism in general describe an impoverished universe devoid of its most important and meaningful aspects. For instance, the Mormon vision includes an infinite regress of gods so that everyone gains meaning by reference to his God, and God by reference to his Super-God etc. So much for Aristotle. Further development of Mormon theology may take it along the road blazed by Cantor and yield a transfinite hierarchy of gods. The lack of consensus concerning the supernatural need be no more significant than the lack of consensus in aesthetics or in science itself. One may anticipate the emergence of religions which are tolerant and open to change and development. They would view our understanding of the supernatural as something which is growing and developing just as is our understanding of the natural.

Of course, the skeptic may, and should, question the validity and significance of religious experiences. He can quite easily accept their subjective reality but deny their ontological significance. He can write them off as hallucination and illusion and declare that God is dead. But, as we have seen, a first class skeptic like Hume can write off practically everything else also. Not only the gods, but also honor, truth, rationality, reality and many other aspects of our universe are at stake. Of course, Hume himself was only showing that one couldn't have "certainty" regarding these or any other aspects of the universe. But absolute certainty isn't required. It's enough to make our choices and live by them. And most people prefer to live in a rich rather than an impoverished universe. The time may be ripe for a non-Euclidean synthesis, of the sciences, the humanities, and theology.

## 9. Conclusions

One of our most powerful myths is the story of the tower of Babel where by an original unity was shattered into diversity. Much of our intellectual history consists of attempts to find common grounds upon which to erect unified sciences. Many bewail the unsatisfactory situation in the social sciences, humanities, and theology where many schools compete for dominance.

This situation contrasts with the seeming harmony of the natural sciences. The claim we have made here is that the relative lack of polemics in the hard sciences is not due to a consensus concerning fundamental theoretical structures, but is instead a much more complicated sociological fact. For instance, a close study of the various electron models reveals tremendous differences. The differences between the Schrodinger, Dirac, and Feynman theories of an electron appear to us as huge as the differences between the Freudian, Skinnerian and Piagetian theories of human behavior. But physicists don't mind the diversity. They take an eclectic approach using whichever model seems most appropriate under the circumstances. By contrast, many psychologists feel that the alternative approaches to human behavior are competing theories and only one of them will eventually prevail.

In studying complex phenomena there are always a variety of possible approaches. Too much of an insistence on consensus results in very impoverished starting points such as those taken by Russell and Carnap . From such starting points it is very difficult to get anywhere. But the other extreme often results in a tower of Babel where each scientist has not only his own theories but his own scientific methodology and logic. To paraphrase Goethe: One can convince one self while viewing a great collection of scientific works that nearly each master had a different way of approaching nature. In our opinion, what is required in science is the same thing that is required in art, namely "taste" and "judgment."

To say this, however, is not to opt for an "anything goes" relativism or radical idealism. On the contrary, we have argued that toleration is necessary in order to save realism, and thus avoid such radical solutions. To the extent that agreement can be reached about goals, there are clear and demonstrable relative advantages between various positions. The determination of these advantages, though, may not be something that is clearly formalizable. On the other hand, in those cases where there is no agreement about goals, we should not expect consensus. In fact, our point has been that in these cases the attempt to enforce consensus carries grave risks.